\documentclass[conference]{IEEEtran}
\IEEEoverridecommandlockouts
\usepackage{cite}
\usepackage{algorithmic}
\usepackage{xcolor}
\usepackage{enumerate}
\usepackage{amsfonts,amsmath,amssymb,amsfonts}
\usepackage{algorithmic}
\usepackage{optidef}
\usepackage{algorithm}
\usepackage{graphicx}
\usepackage{textcomp}
\usepackage{subfigure}
\usepackage[section]{placeins}
\def\BibTeX{{\rm B\kern-.05em{\sc i\kern-.025em b}\kern-.08em
    T\kern-.1667em\lower.7ex\hbox{E}\kern-.125emX}}
\begin{document}
\bibliographystyle{IEEEtran}
\title{Reconfigurable Intelligent Surface Enhanced Device-to-Device Communications}

\author{\IEEEauthorblockN{Zelin Ji,
Zhijin Qin}
\IEEEauthorblockA{Queen Mary University of London, London, UK\\
Email: \{z.ji, z.qin\}@qmul.ac.uk}
}
\maketitle

\begin{abstract}
Reconfigurable intelligent surface (RIS) technology is a promising method to enhance the device-to-device (D2D) communications. To maximize the sum rate of the cellular and D2D networks, a joint optimization of the position and the phase shift of RIS in D2D communications is considered in this paper. To solve the non-convex sum rate maximum problem, we propose a novel convolutional neural network (CNN) based deep Q-network (DQN) that jointly optimizes the RIS position and its phase shift with lower complexity. Numerical results illustrate that the proposed algorithm can achieve higher sum rate compared to the benchmark algorithms, meanwhile meeting the quality of service (QoS) requirements at D2D receivers and the base station (BS).
\end{abstract}


\section{Introduction}
As one of the key technologies of the fifth generation (5G) and beyond communication systems, device-to-device (D2D) communications can enhance the communications performance by reducing the latency and improving energy and spectrum efficiency. However, due to the underlay resource reusing mode, the interference management in D2D communications becomes challenging~\cite{8541128}. The existing works on D2D communications mainly focus on transmit power and channel assignment optimization ~\cite{7579565,7913583}. Unfortunately, due to the fast channel variations in D2D communications, few works contribute to the optimization of wireless communication environment.

Reconfigurable intelligent surfaces (RISs) have attracted extensive attension in wireless communications due to the ability of proactively modifying the wireless communication environment. Equipped with an array of low-cost passive reflecting elements, RISs can adjust the phase shift and amplitude of each element. Compared with conventional relays, the advantages of RIS include energy consumption reduction and achievable data rate improvement~\cite{8796365}. The performance improvements have been verified by an RIS-based wireless communication prototype~\cite{9020088}. The phase of each element is controlled by the RIS controller, which receives control signal from the base station (BS). Although the control signal of RIS can be analog using varactors to provide continuous phase shift~\cite{7510962}, the large response time and low phase accuracy of varactors make it impractical for wireless communications. In~\cite{4476079, 8930608}, theoretical analyses have been provided for multi-bit controlled elements to strike a tradeoff between the performance and complexity. 

However, to benefit the overall system performance, the optimization of RIS becomes challenge due to the large number of reflecting elements~\cite{El_Mossallamy_2020}. Many approaches have been applied to optimize RIS to achieve higher throughput or energy efficiency. Particularly, RIS has been successfully adopted in D2D networks in~\cite{cao2020sum,fu2020reconfigurable,ch2020reconfigurable} to maximize the sum rate in D2D communication systems. To solve the non-convex sum rate maximizing problems, they tend to find the sub-optimal solution by using the block coordinate descent~\cite{cao2020sum} and Riemannian pursuit method~\cite{fu2020reconfigurable}. To achieve performance-complexity tradeoff, the projected sub-gradient method is leveraged for the phase shift design~\cite{ch2020reconfigurable}. However, the channels in D2D networks vary over time, resulting algorithms based on the long-term optimization and high complexity not applicable any more. 

Recently, machine learning (ML) methods, especially deep learning (DL), have become promising tools to address explosive mass data, mathematically intractable nonlinear non-convex problems and high-computation issues. DL based approaches can significantly reduce the complexity, and have been adopted in wireless communication systems, e.g., physical layer communications~\cite{8663966} and resource allocation \cite{8647866}. Motivated by the potential applications of DL in solving sophisticated optimization problems, the authors in \cite{taha2019enabling} have adopted the DL method for designing the RIS reflection matrices with restricted channel state information (CSI). 

As a novel branch of DL, deep Q-network (DQN) has been proposed~\cite{mnih2013playing}. Embraced with reinforcement learning, DQN enables agents to learn and build knowledge by interacting with the environment and maximizing the desired reward, thereby showing its great potential in circumventing challenges of conventional DL. Particularly, DQN is beneficial to discrete phase shift design and wireless communication systems where radio channels vary over time. Leveraging DQN, the authors in~\cite{huang2020reconfigurable} proposed to jointly design the beamforming at the BS and phase shifts at the reflecting RIS to maximize the sum rate.

Note that the applications of DQN in communication systems usually eoploit the fully connected layer instead of convolutional layers, which results in significant increasing on the number of training parameters~\cite{554195}. This motivate us to combine DQN with convolutional neural network (CNN) to jointly optimize the position and phase shift of the RIS, thereby reducing the number of parameters and computation complexity.
The major contributions of this paper are summarized as follows.
\begin{enumerate}[1)]

\item The position and phase shift of RIS are jointly optimized for the D2D networks, which is valuable to be considered before the installation of the RIS.

\item A novel CNN based DQN structure is proposed to reduce the number of training parameters significantly, therefore speed up the convergence time especially when the number of D2D pairs, cellular users and RIS elements are large.

\item In order to meet the quality of service (QoS) requirements of the D2D receivers and the BS, a dynamic reward is defined. 
\end{enumerate}

\section{System Model and Problem Formulation}
We consider the uplink transmission of a D2D network, which includes $K$ cellular users, communicate with the BS in the conventional cellular mode, and $I$ D2D pairs communicate with each other. To enhance transmission performance of the network, an RIS composed of $N$ passive elements is installed. Assuming that the $i$th D2D transmitter $D_{t,i}$ communicates with the corresonding receiver $D_{r,i}$ by reusing the resource block for the uplink of the $k$th cellular user $CU_k$, then $CU_k$ becomes the source of interference for $i$th D2D pairs.
\begin{figure}[t]
\centering
\includegraphics[width=\linewidth,height=0.66\linewidth]{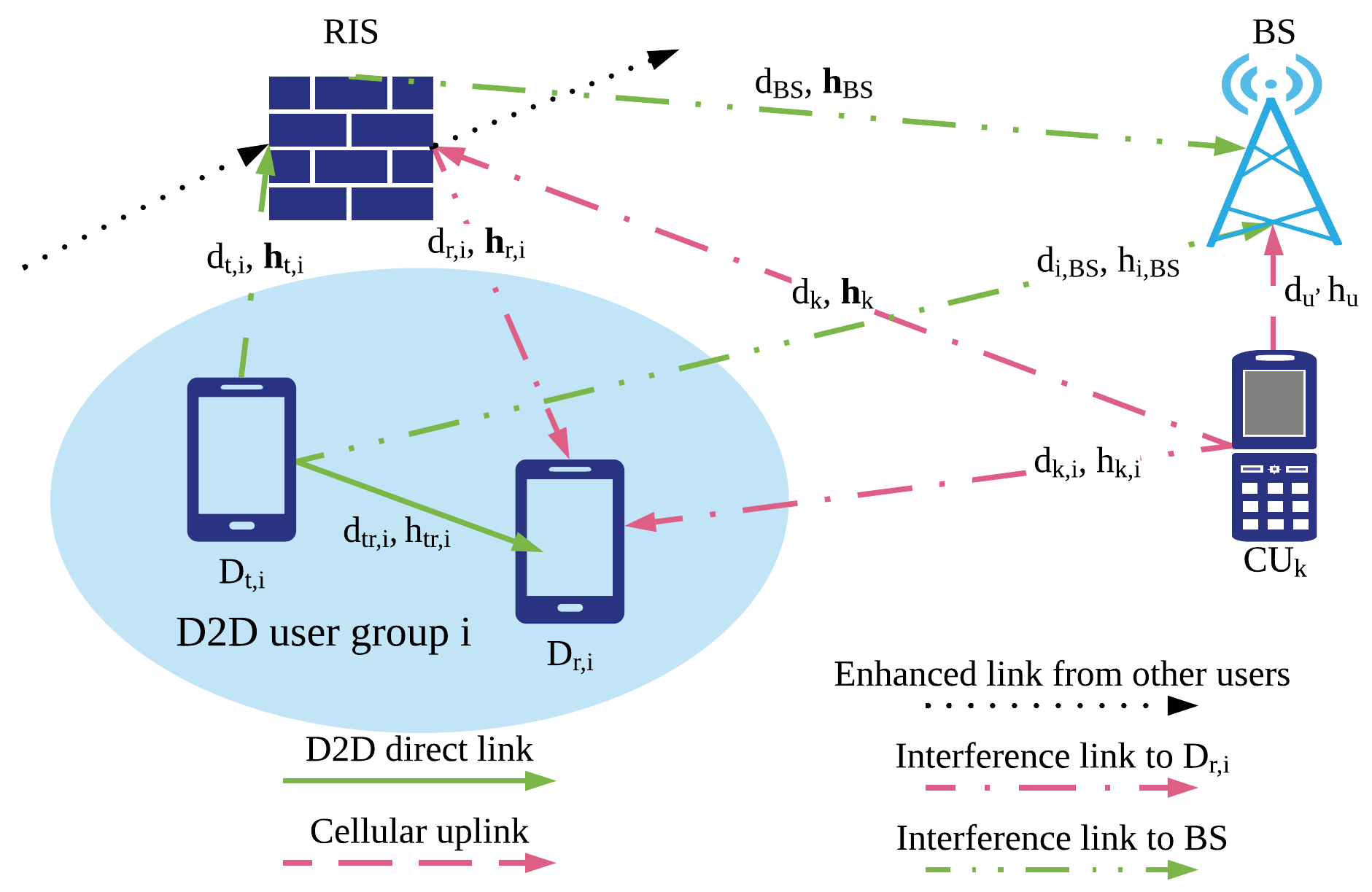}
\centering
\caption{System model of the of RIS enhanced D2D network.}
\label{fig1}
\end{figure}
\subsection{System description}
Practically, the location of a BS is fixed as shown in \ref{fig1}. 
We define the channel gain $\boldsymbol {h}_{t,i}^{\boldsymbol T}$ from the $i$th D2D transmitter $D_{t,i}$ and the RIS as
\begin{equation}
\boldsymbol {h^T_{t,i}} = [h^1_{t,i}, \dots, h^n_{t,i}, \dots, h^N_{t,i}]^{\boldsymbol T},
\label{eq1}
\end{equation}
where $\boldsymbol {h^T_{t,i}}$ represent transpose matrix of $\boldsymbol {h_{t,i}}$, $h^n_{t,i}$ represents the Nakagami fading channel gain between D2D transmitter $D_{t,i}$ and the $n$th RIS element, which can be denoted as
\begin{equation}
h^n_{t,i} = \beta m \sqrt{d_{t,i}^{-\alpha}}, \forall n\in N,
\label{eq2}
\end{equation}
where $\beta$ is a constant value, $m$ is a variable that represents the Nakagami distribution, $d_{t,i}$ is the distance between the $i$th D2D transmitter $D_{t,i}$ and the RIS, and $\alpha$ represents the path loss exponent. Similarly, the channel gain between RIS and the $i$th D2D receiver $D_{r,i}$, between the $k$th cellular user $CU_k$ and RIS, between RIS and BS are defined as $\boldsymbol {h}_{r,i} \in {\mathbb C}^{1\times N}$, $\boldsymbol {h}_k^{\boldsymbol T}\in {\mathbb C}^{N\times1}$ and $\boldsymbol {h}_{BS}\in {\mathbb C}^{1\times N}$, respectively. Note that the channel gains towards the RIS have the dimension ${\mathbb C}^{N\times 1}$, while the gains from the RIS have the dimension ${\mathbb C}^{1\times N}$. The phase shift and amplitude attenuation $A$ for all the RIS elements can be expressed by $\boldsymbol {\Theta} \triangleq diag[A e^{j{\theta_1}}, A e^{j{\theta_2}}, \dots, A e^{j{\theta_N}}]$, where $A \in [0,1]$ and $\theta \in [0,2\pi)$.

For the line-of-sight (LoS) links, we denote $h_{tr,i} = \beta m \sqrt{(d_{tr,i})^{-\alpha}}$ as the channel gain for the link between D2D pairs ($D_{t,i}$ and $D_{r,i}$), where $d_{tr,i}$ represents the distance between $D_{t,i}$ and $D_{r,i}$. The channel gain from the $k$th cellular user $CU_k$ to the $i$th D2D receiver $D_{r,i}$, from the $i$th D2D transmitter $D_{t,i}$ to BS, and from the $k$th cellular user $CU_k$ to BS can be denoted by $h_{k,i}$, $h_{i,BS}$, and $h_u$ in the similar way. 

Overall, the channel gain $h_D$ between D2D link and $h_{C,D}$ between cellular user $CU_k$ to D2D receiver $D_{r,i}$ can be given by

\begin{equation}
h_D = \underbrace{\boldsymbol {h}_{t,i}^{\boldsymbol T} \boldsymbol {\Theta} \boldsymbol {h}_{r,i}}_{\text{Reflection link}}  + \underbrace{h_{tr,i}}_{\text{LoS link}},
\end{equation}
and
\begin{equation}
h_{C,D} = \boldsymbol {h}_{r,i} \boldsymbol {\Theta} \boldsymbol {h}_k^{\boldsymbol T} + h_{k,i}.
\end{equation}
In the similar way, the overall channel gain from $CU_k$ and $D_{t,i}$ to the BS can be represented as $h_C$ and $h_{D,BS}$, respectively.
The signal $y_i$ received by $D_{r,i}$ is
\begin{equation}
y_i = h_D x_i + \underbrace{h_{C,D} x_k}_{\text{Interference signal}} + \underbrace{z}_{\text{Noise}},
\label{eq4}
\end{equation}
where $x_{t,i} \triangleq \sqrt{p_i} u_i$ and $x_k \triangleq \sqrt{p_k} u_k$ denotes the signal from $D_{t,i}$ and $CU_k$, $p_i$ and $u_i$ denotes transmission power of the D2D transmitter and unit variance entries with zero mean, $z\thicksim N(0,\sigma^2)$ denotes the AWGN noise signal with $\sigma^2$ variance. Then, the signal-to-interference-plus-noise ratio (SINR) $\gamma_i^D$ at $i$th D2D receiver and $\gamma_k^C$ at the BS for the $k$th cellular user can be denoted as
\begin{equation}
\gamma_i^D = \frac{p_i |h_D|^2}{\sum^K_{k=1} \rho_{k,i} p_{k} |h_{C,D}|^2+\sigma^2}, \end{equation}
and
\begin{equation}
\gamma_k^C = \frac{p_k |h_C|^2}{\sum^I_{i=1} \rho_{k,i}p_{i}|h_{D,BS}|^2+\sigma^2},
\label{eq5}
\end{equation}
where $\rho_{k,i}$ is the resource reuse coefficient of $k$th cellular user and $i$th D2D pair, $\rho_{k,i}=1$ when $i$th D2D pair reuses the resource of $CU_k$, and $\rho_{k,i}=0$ otherwise.

\subsection{Problem formulation}
Our objective is to maximize the sum rate of D2D and cellular networks, which could given by
\begin{equation}
\begin{aligned}
R&=\sum \limits_{i=1} R_i +\sum \limits_{k=1} R_k \\
&=\sum \limits_ {i=1} B_i log_2(1+\gamma_i^D)+\sum \limits_ {i=k} B_k log_2(1+\gamma^C_k),    
\end{aligned}
\label{eq6}
\end{equation}
where $B_i$ and $B_k$ represents the bandwidth of $i$th D2D transmission link and the uplink of $k$th cellular user $CU_k$. 

This paper aims to maximize the sum rate in (\ref{eq6}) by jointly optimize the phase shift and position of RIS. The joint RIS positioning and phase shift problem can be formulated as
\begin{maxi!}|l|
{\{\boldsymbol {\Theta},\boldsymbol {s_{RIS}},\boldsymbol \rho\}}{R}
{\label{eq7}}{\text{P1:}}
\addConstraint{\gamma_i^D \geq \gamma_{min}^D, \forall i\label{objective:c1} }
\addConstraint{\gamma_k^C \geq \gamma_{min}^C, \forall k\label{objective:c2} }
\addConstraint{0 < \theta_n \leq \pi, \forall n \in N\label{objective:c3} }
\addConstraint{\boldsymbol{s_{RIS}}\in \mathbb{R}^2, \label{objective:c6} }
\end{maxi!}
where $\gamma_{min}^D$ and $\gamma_{min}^C$ are the minimum SINR requirements at the D2D receiver and the BS, respectively. Coordinate $\boldsymbol{s_{RIS}}$ restricts a 2-dimension space for the installation of the RIS. The constraint (\ref{objective:c3}) makes $P1$ non-convex. To solve the non-convex problem by mathematical tools, we have to utilize exhaustive search, which is impractical when the number of D2D pairs, cellular users, and RIS elements become large. Generally, classical mathematical tools can be leveraged to acquire suboptimal solutions~\cite{cao2020sum,fu2020reconfigurable,ch2020reconfigurable}. Alternatively, instead of solving challenging non-convex problem by mathematical tools, we adopt CNN based DQN algorithm, which is more applicable to solve problems with high dimension inputs and large state and action space.


\section{Problem Solutions}
In this section, we propose a CNN based DQN learning model is proposed to solve the joint RIS positioning and phase shift problem. Particularly, the RL components are first defined and the DQN components are then introduced. Then the proposed algorithm is explained in detail. 
\subsection{RL components definition}
In our proposed algorithm, the state contains the position information of D2D users, cellular users, BS and RIS, as well as the phase shift $\boldsymbol \Theta$. The position vectors $\boldsymbol{S_t}=[\boldsymbol{s_{t,1}},\dots,\boldsymbol{s_{t,I}}]$, $\boldsymbol{S_r}=[\boldsymbol{s_{r,1}},\dots,\boldsymbol{s_{r,I}}]$, $\boldsymbol{S_u}=[\boldsymbol{s_{u,1}},\dots,\boldsymbol{s_{u,K}}]$, $\boldsymbol{s_{RIS}}$ and $\boldsymbol{s_{BS}}$ represents the position information of D2D transmitters, D2D receivers, cellular users, RIS and BS, respectively. The input state $\cal {S} = [\boldsymbol{S_t},\boldsymbol{S_r},\boldsymbol{S_u},\boldsymbol{s_{RIS}},\boldsymbol{s_{BS}};\boldsymbol \Theta]$, which has a cardinality $|\cal S|$ of ($2I + K + N + 2$).

Action set $\cal A$ represents the possible action choice for the RIS controller. Generally, the position of RIS are fixed after installation, while the phase shift can be adjusted, so the action space contains the phase shift adjustment and position choice of RIS. At iteration $t$, action $a_t$ consists of two parts: \romannumeral1) the variable quantity of phase shift matrix, $\Delta \boldsymbol \Theta = \{\Delta \theta_1, \dots, \Delta \theta_N\}$, where $\Delta \theta_n \in \{-\delta, 0, +\delta \}, \forall n \in N$; \romannumeral2) the position choice of RIS, $\boldsymbol s \in \{v_1, \dots, v_O\}$, where $O$ represents the number of grids in the communications system. Formally, the action $a_t = [\Delta \boldsymbol \Theta; \boldsymbol {s_{RIS}}]$, which has a cardinality $|a|$ of ($N + 1$). Action set $\cal A$ includes all possible actions with the cardinality $|{\cal A}| = 3^N\times O$.

The reward represents whether we encourage or punish an action, so it is defined based on the objective function given in (\ref{eq6}). For a successful transmission at iteration $t$, i.e., the constraints (\ref{objective:c1}) and (\ref{objective:c2}) are satisfied, the reward $r_s$ can be defined as $r_s = R(t)$,
where $R(t)$ represents the achievable rate $R$ at iteration $t$. However, if any of the constraints are not satisfied, the expected QoS cannot be achieved. This kind of action results in penalty due to energy waste, and we defined the new reward for the transmission failure as
\begin{equation}
r_\text{f} = 
\begin{cases}
\sum \limits_{i=1} R_i,&\text{if (\ref{objective:c1}) is not satisfied};\\
\sum \limits_{k=1} R_k,&\text{if (\ref{objective:c2}) is not satisfied};\\
0,&\text{otherwise};
\end{cases}
\label{eq9}    
\end{equation}
The fail reward is to encourage the communication system to improve the SINR which is not satisfied the requirement.
The overall reward can be expressed as
\begin{equation}
r_t=
\begin{cases}
r_s,&\text{if (\ref{objective:c1}) and (\ref{objective:c2}) are satisfied};\\
r_\text{f},&\text{else}.
\end{cases}
\label{eq10}
\end{equation}
The optimal action-value function obeys an important identity known as the Bellman equation. The optimal strategy is to select the action that maximizes~\cite{mnih2013playing}:
\begin{equation}
Q^*(s_t, a_t)=\mathbb{E}_{s_{t+1}}[r_t+\Gamma\max\limits_{a'\in A} Q^*(s_{t+1},a')|s_t,a_t],
\label{eq11}
\end{equation}
where $Q^*(s, a)$ is the desired value function such that $Q(s_t, a_t) \rightarrow Q^*(s_t, a_t)$ as $t \rightarrow \infty$. However, it is impractical since the iteration is discrete. Instead, the neural networks (NN) are applied to be function approximator to estimate the action-value function, i.e., $Q(s_t, a_t; \boldsymbol W) \approx Q^*(s, a)$. When the state and action space become large, this method does not need to maintain the large Q-table as conventional RL approaches do, thereby expanding the applications of RL in wireless communications greatly.
\begin{figure}[t]
\centering
\includegraphics[width=\columnwidth, height =0.62\columnwidth]{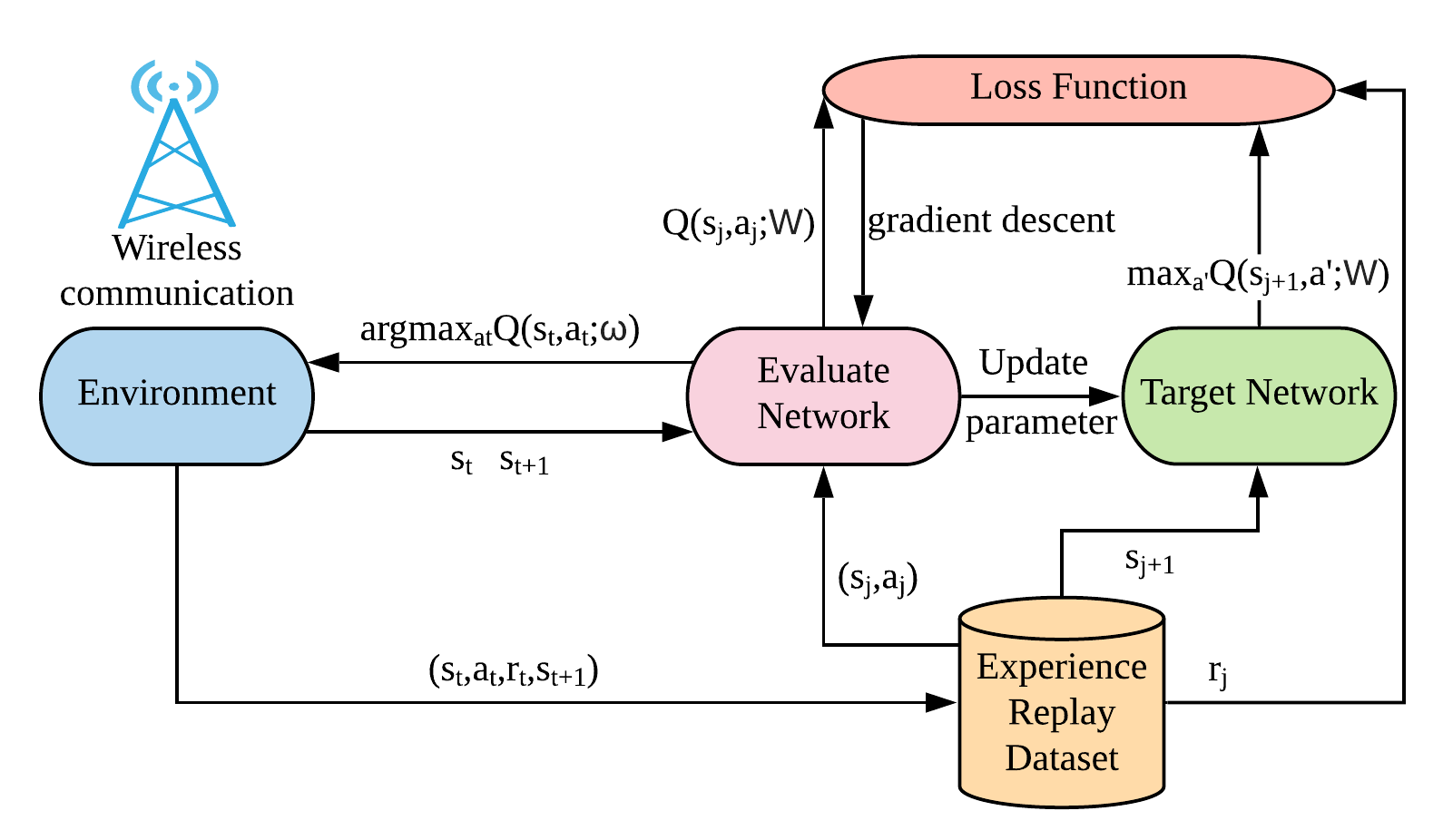}
\centering
\caption{The proposed CNN based DQN algorithm for the RIS position and phase shift optimization.}
\label{fig3}
\end{figure}
\begin{table*}[ht]
\begin{center}
\caption{Position and phase shift input format to CNN based DQN}
\begin{tabular}{|c|c|c|c|c|c|c|c|c|c|c|}
\hline
\textbf {Position} &
$\boldsymbol s_{t,1}$ &
$\boldsymbol s_{r,1}$ &
\dots &
$\boldsymbol s_{t,I}$&
$\boldsymbol s_{r,I}$&
$\boldsymbol s_{u,1}$&
\dots&
$\boldsymbol s_{u,K}$&
$\boldsymbol s_{RIS}$&
$\boldsymbol s_{BS}$\\ \hline
\textbf{Phase} &
$\theta_1$&
$\theta_2$&
\dots&
$\theta_n$&
\dots&
$\theta_N$&
0&
0&
0&
\dots\\
\hline
\end{tabular}
\label{tab1}
\end{center}
\end{table*}

\subsection{DQN components}
Leveraging the NN, the DQN model can find the relationship between the input location of D2D and cellular users and the corresponding position and phase shift combination of RIS. The components in DQN is defined as
\begin{itemize}
\item Agent: The agent in our DQN model is RIS controller. RIS controller will proceed the inputs and execute the outputs of DQN to control RIS.
\item Input: The DQN model takes the states $\cal S$ as the input, which includes the position and phase shift information.
\item Output: The output of the DQN model is the evaluate Q-value for state-action pairs. The output layer contains $|{\cal A}|$ units, which represents the number of possible actions. As shown in Fig. \ref{fig3}, two identical networks are set: evaluation network and target network. In the evaluation network, current state $s_j$ is the input information, and the output is the evaluate Q-value for each action. In the target network, next expected state $s_{j+1}$ is the input, while the output is the Q-value for the each action in the next state.
\end{itemize}

\subsection{DQN learning process}
At each iteration, the BS observes the states and send them to RIS controller as the input. DQN in RIS controller will process the input and output the approximately Q-value for each state-action pair. The policy for selecting actions is to make a trade-off between exploitation and exploration, so we apply decaying $\epsilon$-greedy algorithm \cite{liu2020ris}. The agent will choose actions uniformly from ${\cal A}$ with probability of (1-$\epsilon$), while choosing the action which maximize the Q-value with the probability of $\epsilon$. 

In the adopted DQN, the training data set, also named replay memory ${\cal D}=[\boldsymbol e_1,\dots, \boldsymbol e_t, \dots]$ for NN is stored according to agent's experience at each iteration $t$, where the experience $\boldsymbol e_t=(s_t,a_t,r_t,s_{t+1})$ is called transition, including the state, action and reward information. The training minibatch $(s_j,a_j,r_j,s_{j+1})$ is sampled from the training data set. During the training process, parameters are updated to the Q estimation network at each step to generate the estimated Q-value. Q target network is updated after every $g$ steps according to the parameters in the Q estimation network. The training process for DQN is to minimize the error function which represents the estimated Q-value and the realistic Q-value. For the DQN in this work, the error function can be expressed by:
\begin{equation}
\text{Loss}(\boldsymbol W)=\mathbb E[(q_{target}-Q(s_j,a_j;\boldsymbol{W}))^2],
\label{eq13}
\end{equation}
where $q_{target}=r_j+\Gamma \max_{a'}Q(s_{j+1},a';\boldsymbol W^*)$ is the target Q-value for minibatch $j$, which is the output of Q target network. $\boldsymbol W$ and $\boldsymbol W^*$ denotes the weights of the evaluation network and the target network, respectively. The weights are optimized by the gradient descent method~\cite{mnih2013playing}. 

\subsection{Proposed CNN based DQN algorithm for the control of RIS}
Generally, in each iteration, the fully connected network need to train $ |{\cal S}| \times |{\cal A}|$ parameters, and even more if there are hidden layers. In order to reduce the number of training parameters and improve the training complexity, CNN is an effective approach. The proposed CNN algorithm consists of a convolutional layer, followed by a flatten layer and a fully connected layer which connected to the output layer. 
\begin{itemize}
    \item Convolutional layer: In the proposed CNN based DQN algorithm, the convolutional layer is employed to extract the information of input, i.e., the position and the phase shift information in this paper. The inputs information are stored in a matrix shown in Table~\ref{tab1}. The number of columns $U = \max (2I+K+2,N)$. To make the format complete, the rest elements are set to 0. Mathematically, the input of the convolutional layer is ${\cal S} \in \mathbb R^{2 \times U}$, then the output of convolutional layer is given by:
    \begin{equation}
    {\boldsymbol v}^\lambda_1 = f({\cal S} \otimes \boldsymbol \omega^\lambda_1 + \boldsymbol b^\lambda_1),
    \label{eq12}
    \end{equation}
    where $f(\cdot)=\max(0,\cdot)$ is rectified linear unit (ReLU) activation function, $\otimes$ represents the convolution operation, $\boldsymbol \omega^\lambda_1 \in \mathbb R^{2 \times \mu}$ and $\boldsymbol b^\lambda_1$ represents the convolution kernel and bias in first convolutional layer of the feature map $\lambda$. Note that output feature map is ${\boldsymbol v}^\lambda_1 \in \mathbb R^{U-\mu+1}$.
    \item Flatten layer: A flatten layer is applied after convolutional layer, whose inputs are $\Lambda$ feature maps extracted by the convolutional layer, and it generates the feature vector $\boldsymbol v_2 \in \mathbb R^{\nu}$, where $\nu=\Lambda \times (U-\mu+1)$.  
    \item Hidden layer: The hidden layer is fully-connected with 256 rectifier units.
    \item Output layer: The output layer is a fully-connected linear layer with single output of each valid action, represents the evaluated Q-value or the target Q-value for a specific state action pair in evaluation network or target network, respectively. 
    \item Complexity comparison: Compared with the conventional DQN need to train $(|{\cal S}|\times|{\cal A}|)$, the number of training parameter in the first layer of CNN based DQN is only $(2\times \mu \times \Lambda)$. Clearly, CNN can reduce the complexity as state and action space becomes large. 
\end{itemize}
\begin{algorithm}
\caption{CNN based DQN algorithm for the RIS}
\label{algo1}
\begin{algorithmic}[1]
\STATE \textbf{Input}: Environment simulator, Q network, replay memory $\cal D$, minibatch size;\
\STATE Initialize: action-value function Q with random weights $\boldsymbol W$, replay memory $\cal D$, RIS position and phase;\
\REPEAT
\STATE for each iteration step:
\STATE choose action $a_t$ from aciton space $\cal A$ according to \\$\epsilon$-greedy algorithm;\
\STATE Execute $a_t$, calculate dynamic reward $r_t$ by (\ref{eq10}) and observe $s_{t+1}$;\
\STATE Store transition $(s_t,a_t,r_t,s_{t+1})$ in $\cal D$;\
\STATE Replay memory:\
\STATE Sample random minibatch of transitions\\ $(s_j,a_j,r_j,s_{j+1})$ in $\cal D$;\
\STATE Calculate $q_{target}$ by (\ref{eq13})\
\STATE Perform a gradient descent step on \\ $(q_{target}-Q(s_j,a_j;\boldsymbol{W}))^2$;\
\UNTIL $s_t$ is the goal state\
\STATE \textbf{Return}: Action value function and optimized action $a$.
\end{algorithmic}
\end{algorithm}

Overall, by receiving the input information of position information of D2D pairs, cellular users and the BS from the wireless environment, the RIS controller can train the weights and update NNs to estimate the action value function. The proposed algorithm learns the policy to jointly optimize the installation position and phase shift of the RIS with much lower complexity. The details of the proposed algorithm are shown in Algorithm \ref{algo1}.

\section{Numerical Results}
\addtolength{\topmargin}{0.01 in}
In this section, performance of the proposed CNN based DQN algorithm is evaluated by comparing it with the benchmark algorithms. We take 1 D2D pair and 1 cellular user in the considered network as an example, i.e., $I=K=1$. We place them in a square with size of $100m \times 100m$. Particularly, D2D transmitter, D2D receiver, cellular user, and the BS are fixed at position (40,20), (60,20), (25,55) and (75,55) respectively. The whole area is divided in $O=25$ identical squares, where RIS can be installed  in any of them. The variable quantity $\delta$ of the phase shift is $\frac{\pi}{4}$. To guarantee the QoS, the minimum SINR requirements for D2D receiver $\gamma^D_{min}$ and for cellular uplink $\gamma^C_{min}$ are -10dB and -13dB, respectively. Gaussian noise variance $\sigma^2$ is -116dBm. The path loss parameter $d$ is set to 3. The learning rate $\alpha$ is set $\frac{1}{\text{Iterations}}$ to guarantee the convergence performance of the algorithm. Discount factor $\Gamma$ is set to 0.9. The exploration rate $\epsilon$ is set to $0$ at the beginning and increases by $0.01$ at each iteration until it reaches $0.9$ so that RL can explore the new actions with reasonable probability while guaranteeing the system performance.




\subsection{Convergence Performance of Proposed Algorithms}
\begin{figure}[t]
\centering
\includegraphics[width=0.9\columnwidth]{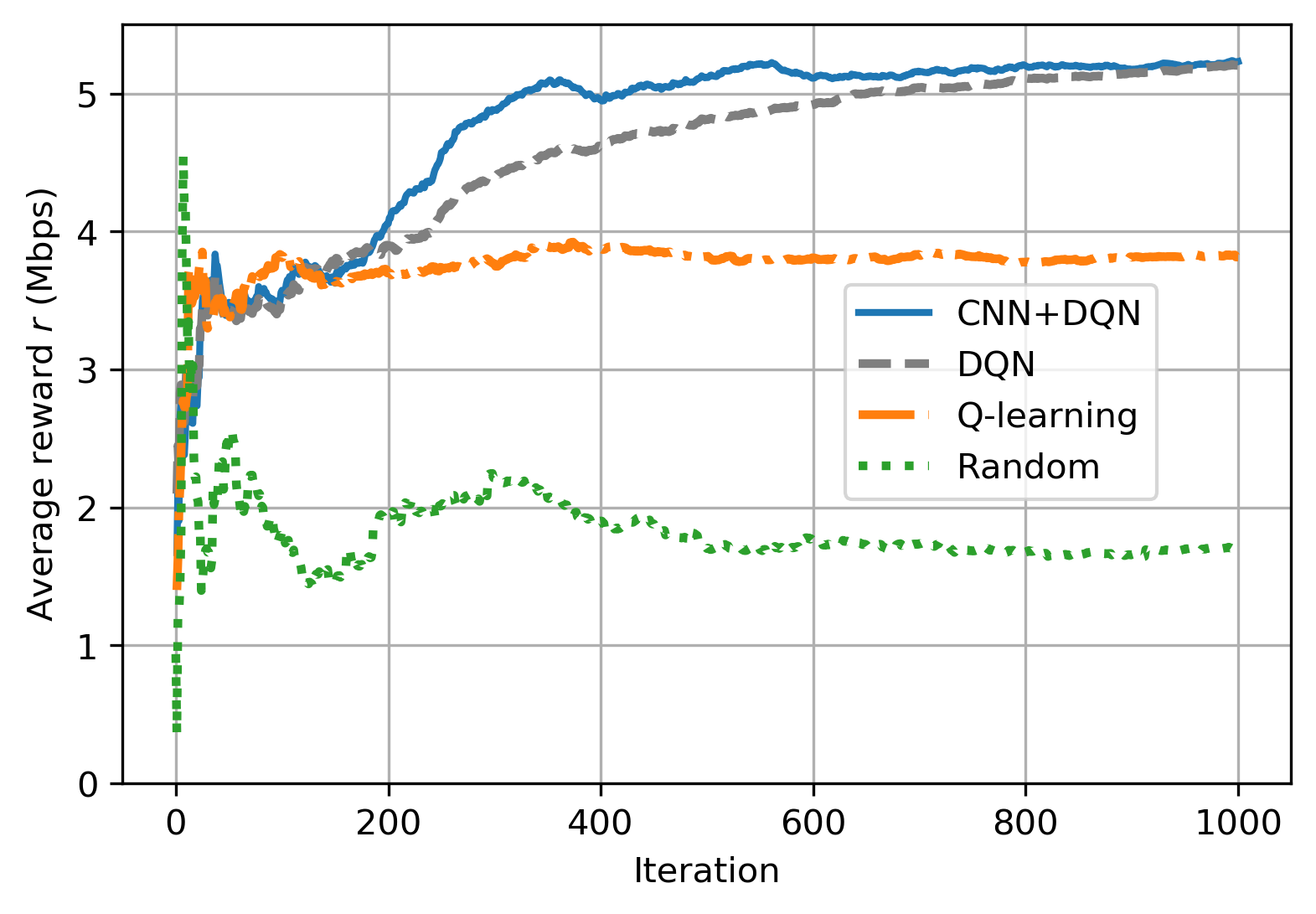}
\centering
\caption{Reward comparison of proposed CNN based DQN algorithm, conventional DQN, Q-learning and Random. $N$ = 16, $p_i$ = 15dBm, $p_k$ = 30dBm, average reward $r = \sum r_t/t$.}
\label{fig5}
\end{figure}

Fig. \ref{fig5} demonstrates the convergence of the proposed CNN based DQN algorithm, while other algorithms including conventional DQN algorithm, and Q-learning algorithm are also demonstrated as benchmarks. In the random benchmark scheme, the actions are chosen randomly from the action space $\cal A$ and the agents do not learn anything from the environment. Note that the reward $r_t$ is calculated by (\ref{eq10}), so it is not equal to the sum rate $R$. From the figure, the proposed algorithm achieves the highest reward with fast convergence speed, while conventional DQN algorithm can achieve the same reward with more iterations. Q-learning can achieve fastest convergence, however, the reward is lower than DQN and the proposed algorithm. Compared with Q-learning, DQN and the proposed CNN based DQN algorithm discard the Q-table, which contains a large number of state-action pairs. Compared with conventional DQN, our proposed algorithm invokes CNN to reduce the number of training parameters. Additionally, the proposed algorithm outperforms the conventional DQN by invoking decaying $\epsilon$-greedy policy. 
\begin{figure}[ht]
\centering
\subfigure[Sum rate over D2D transmit power. $p_k$ = 30dBm.]{
\begin{minipage}[h]{\linewidth}
\centering
\includegraphics[width=0.9\linewidth]{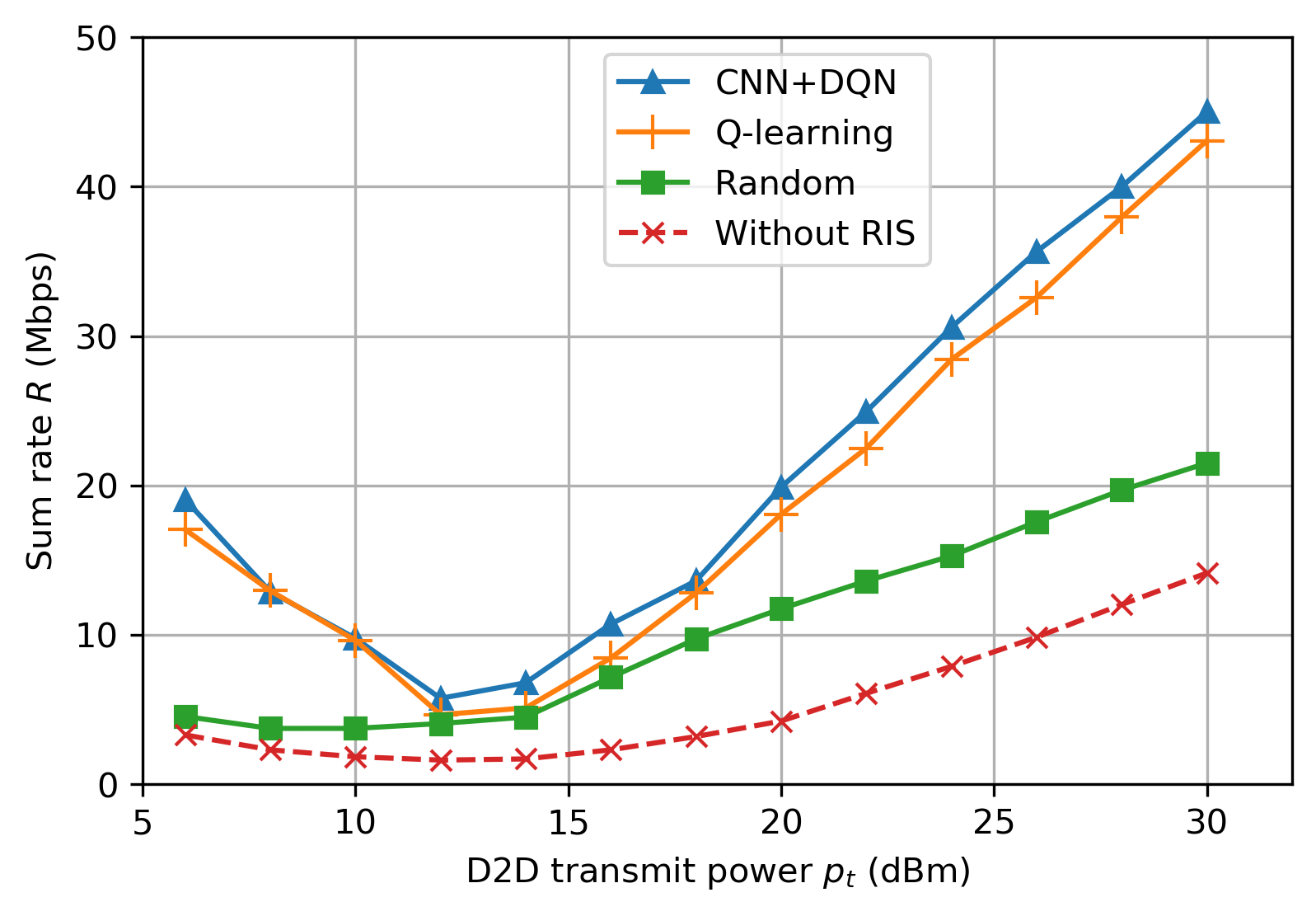}
\end{minipage}%
}%

\subfigure[Sum rate over cellular transmit power. $p_i$ = 15dBm.]{
\begin{minipage}[ht]{\linewidth}
\centering
\includegraphics[width=0.9\linewidth]{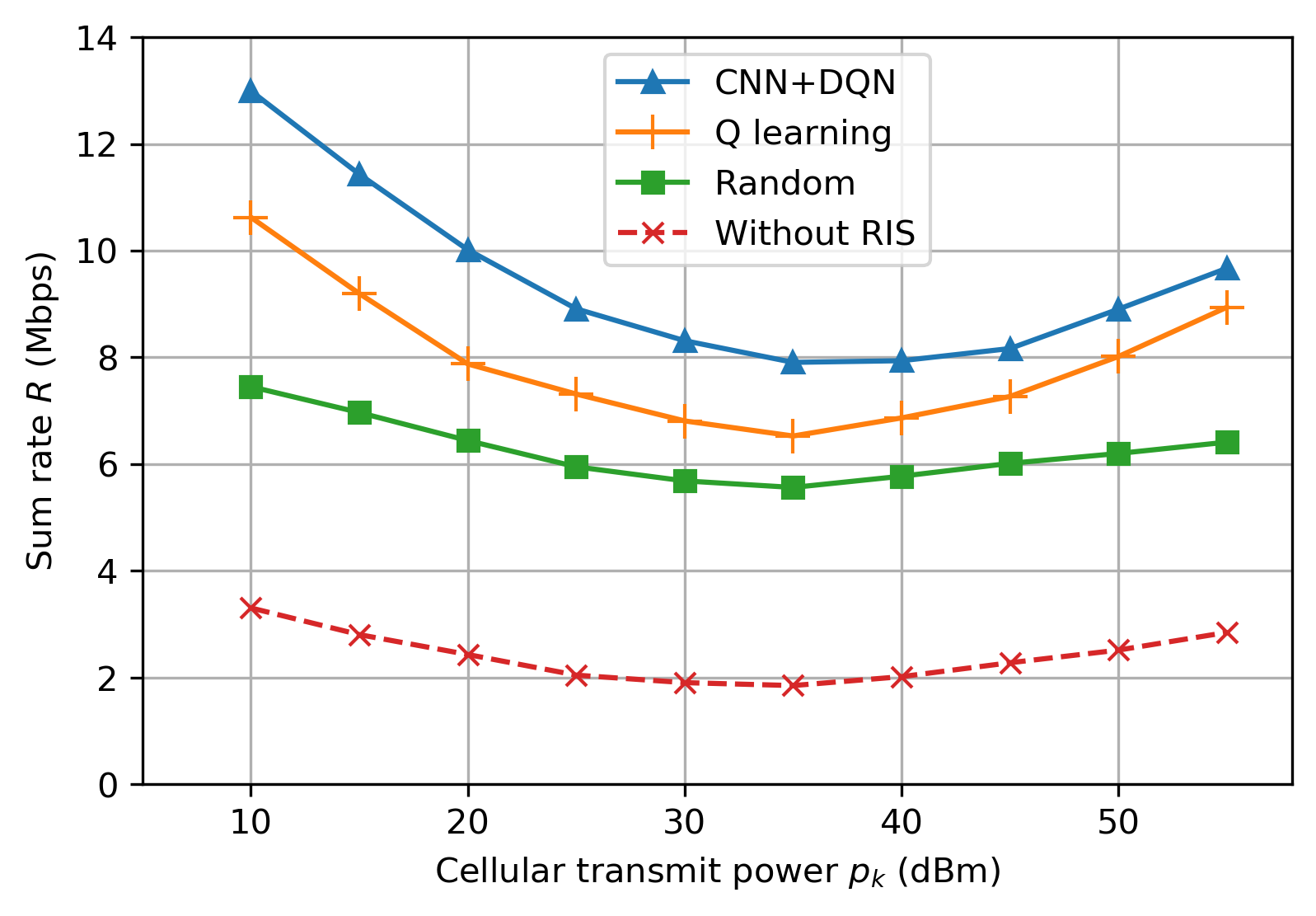}
\end{minipage}%
}%
\centering
\caption{{Sum rate over D2D and cellular transmit power. $N$ = 16.}}
\label{fig6}
\end{figure}

\subsection{Sum Rate versus Transmit Power}
Fig. \ref{fig6}(a) demonstrates the sum rate with different D2D transmit power. We can note that the schemes with RIS always outperform that without RIS in terms of achieved sum rate, which demonstrates the effectiveness of RIS. As the D2D transmit power is relatively low, the sum rate is mainly contributed by cellular links. With D2D transmit power increasing from 5dBm to 12dBm, the sum rate $R$ decreases as the interference from D2D transmitters to the BS becomes severer. When D2D transmit power becomes more than 14dBm, the sum rate $R$ increases with higher D2D transmit power as D2D links contribute more to the sum rate. It is also noted that our proposed CNN based DQN algorithm outperforms the Q-learning and random scheme in all the cases. 

In Fig. \ref{fig6}(b), the sum rate versus transmit power of the cellular user shows the similar trend. The sum rate reach the bottom when the cellular transmit power is 35dBm, which means the both of the D2D receiver and the BS experience severe interference. As the transmit power of the cellular user is beyond 35dBm, the sum rate turns to increase with the higher transmit power from the cellular user.

\subsection{Impact of the number of RIS elements}
\begin{figure}[t]
\centering
\includegraphics[width=0.9\columnwidth]{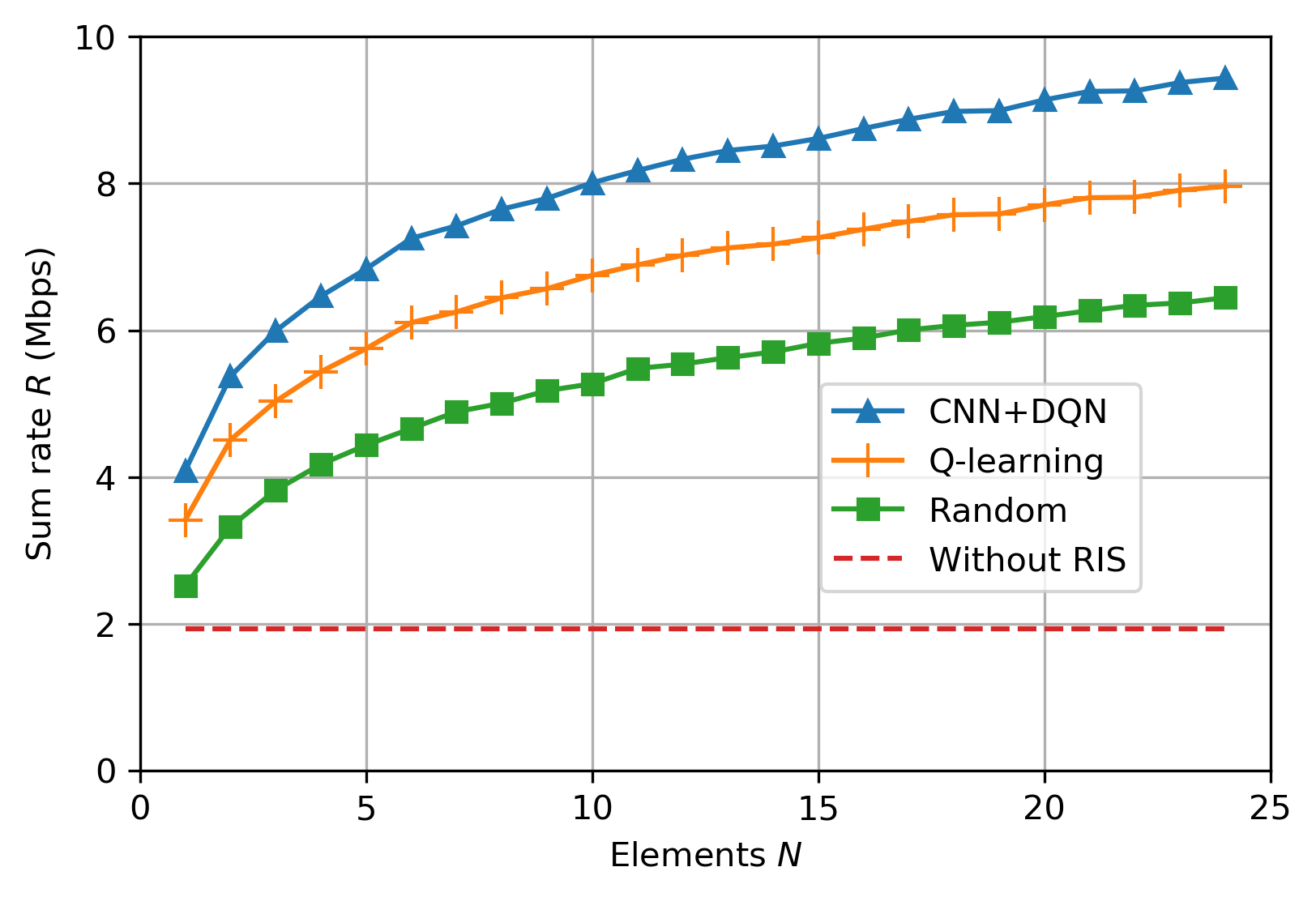}
\caption{Sum rate over number of elements. $p_k$ = 30dBm, $p_i$ = 15dBm.}
\label{fig7}
\end{figure}

Fig. \ref{fig7} demonstrates the impact of the number of RIS elements on the sum rate. In the random phase shift, we optimize the position of RIS, while randomly adjusting the phase shift. It is observed that the sum rate of the considered network increase with more number of RIS elements implemented. This is because the sum rate increases due to the improved gain. However, the performance improvement slows down as the number of elements becomes larger, which is caused by the server interference.

\section{Conclusion}
In this paper, we consider the joint optimization of position and phase shift to maximize the sum rate of the D2D network. To solve the non-convex problem, a novel CNN-combined DQN was proposed to reduce the number of weights to be trained. By interacting with wireless communication environment and receiving real-time feedback, the RIS controller was to be able to learn an policy to optimize the installation position and phase shift of the RIS. The simulation results demonstrated that the proposed algorithm outperforms the benchmark schemes.

\bibliography{Reference}

\end{document}